\documentclass[aps,pra,twocolumn]{revtex4-1}
\usepackage{graphicx}
\usepackage{amsmath}
\usepackage{bm}

\begin{document}
\title{Atomic properties of actinide ions with particle-hole configurations}

\author{M. S. Safronova$^{1,2}$, U. I. Safronova$^3$, and M. G. Kozlov$^{4,5}$}

\affiliation{$^{1}$Department of Physics and Astronomy, University
of Delaware, Newark, Delaware 19716, USA}

\affiliation{$^{2}$Joint Quantum Institute, National Institute of
Standards and Technology and the University of Maryland,
College Park, Maryland 20742, USA}

\affiliation{$^3$University of Nevada, Reno, Nevada 89557, USA}

\affiliation{$^4$Petersburg Nuclear Physics Institute of NRC
``Kurchatov Institute'', Gatchina 188300, Russia}

\affiliation{$^5$St.~Petersburg Electrotechnical University
``LETI'', Prof. Popov Str. 5, St.~Petersburg, 197376, Russia}

\begin{abstract}
We study the effects of higher-order electronic correlations in the
systems with  particle-hole excited  states using a relativistic
hybrid method that combines configuration interaction and linearized
coupled-cluster approaches. We find the configuration interaction
part of the calculation sufficiently complete for eight electrons
while maintaining good quality of the effective coupled-cluster
potential for the core. Excellent agreement with experiment was
demonstrated  for a test case of La$^{3+}$. We apply our method for
homologue actinide ions Th$^{4+}$ and U$^{6+}$ which are of
experimental interest due to a puzzle associated with the resonant
excitation Stark ionization spectroscopy (RESIS) method. These ions
are also of interest to actinide chemistry and this is the first
precision calculation of their atomic properties.
\end{abstract}
\maketitle

\section{Introduction}

Last two decades of progress in atomic, molecular, and optical (AMO)
physics brought forth a plethora of new AMO applications, ranging
from quantum information \cite{ZhaPagHes17} to dark matter searches
\cite{Arv15,RobBleDai17,FJKL17}. These advances required further
development of high-precision theory, first for alkali-metal and
alkaline earth metal atoms, and then for more complicated systems
with larger number of valence electrons. A hybrid method  which
combines configuration interaction and linearized coupled-cluster
approaches was developed  for these purposes
\cite{Koz04,SafKozJoh09,PKST16} and applied for a wide range of
problems in ultracold atoms
\cite{SafKozCla11,SafPorCla12,ZhaBisBro14} and search for physics
beyond the Standard Model of particles and interactions
\cite{PorSafKoz12,SafDzuFla14,SafBudDem17}. This method has been
tested and demonstrated to give accurate results up to four valence
electrons \cite{SafSafCla14,DzuSafSaf14}. Recently, there was much
interest in application of atoms and ions with even more complicated
atomic structure, including lanthanides, actinides, various highly
charged ions and negative ions
\cite{SafDzuFla14,JorCerFri15,TilLeeBou15,TanSykBur16}. In
particular, the ability to treat hole-particle states with good
precisions is needed \cite{BerDzuFla11,HunSanLip16}. The problem of
applying CI+all-order  method  to predict properties of more
complicated systems lies in the exponential scaling of the number of
possible configurations with the number of valence electrons.
However, if the most important sets of configuration are identified
this method may still yield accurate values for larger number of
electrons. In this work, we demonstrate a first accurate calculation
of the systems with 8 valence electrons  using the all-order
effective Hamiltonian combined with a large-scale configuration
interaction calculation in  a valence sector.

We demonstrate the methodology on the  example of Th$^{4+}$ and
U$^{6+}$. These ions are of particular interest to actinide chemistry, as
U and Th usually occur in chemical compounds and solutions as
multiply-charged cations, most commonly near the Rn-like ion with a
closed shell configuration \cite{MorEdeFug11}. No spectroscopy data
exists for excited levels of the Rn-like U and Th ions, i.e. there
is no experimental data for any of the energy levels. A very
successful program was established to measure the dipole and
quadrupole polarizabilities of Th ions with resonant excitation
Stark ionization spectroscopy (RESIS) method
\cite{HanKeeLun10,KeeLunFeh11,KeeHanWoo11,KeeSmiLun12,KeeSmiLun13}.
In the RESIS method,  a non-penetrating Rydberg electron is attached
to the ion of interest to measure the binding energies of the
resulting high-L Rydberg states \cite{HanKeeLun10}. The energy
levels in the fine structure pattern are determined by the
properties of the core ion, mainly by its dipole and quadrupole
polarizabilities. Therefore, these properties can be extracted from
the Rydberg high-L energy measurements. This method was successful
for Th$^{4+}$ and Th$^{3+}$, but failed completely for the U$^{6+}$
ions - no resolved spectral features were observed
\cite{private,Sco15}. This is particulary puzzling since Th$^{4+}$
and U$^{6+}$ were predicted to have very similar energy levels
structures \cite{SafSaf11}. However, the theory calculations were
not  of sufficient precision to definitively establish the order of
the first two excited levels. Differences in the properties of the
low-lying metastable  states of Th$^{4+}$ and U$^{6+}$ may provide
an explanation for the failure of the RESIS experiments in U$^{6+}$
\cite{private,Sco15}. In summary, reliable  precision calculations
are needed to resolve this puzzle.

\begin{table*} [ht]
\caption{\label{tab1} Energies of Xe-like lanthanum, La$^{3+}$. The
CI+all-order results obtained considering $5s^2$ to be a core shell
are listed in rows labeled ``6-el''. The CI+all-order results
obtained considering $5s^2$ to be valence electrons are listed in
rows labeled ``8-el''. Results obtains with small, medium and large
sets of configurations are given in the corresponding rows.  The
results are compared with experimental data complied in NIST
database \cite{NIST}.}
\begin{ruledtabular}
\begin{tabular}{lrrrrrrr}
\multicolumn{1}{c}{Level} &
\multicolumn{1}{c}{COWAN} &
\multicolumn{1}{c}{6el-small} &
\multicolumn{1}{c}{6el-large} &
\multicolumn{1}{c}{8el-small} &
\multicolumn{1}{c}{8el-medium} &
\multicolumn{1}{c}{8el-large} &
\multicolumn{1}{c}{NIST}\\
\hline \\[-0.8pc]
 $5p^6  \ ^1S_0$&    0    &     0 &      0&       0&        0&      0  &   0       \\
 $5p^54f\ ^3D_1$&  150023 & 138453& 137939&  143780&   142476& 142168  &  143354.7\\
 $5p^54f\ ^3D_2$&  152843 & 141183& 140609&  146591&   145219& 144910  &  145949.0\\
 $5p^54f\ ^3G_5$&  155822 & 144530& 143879&  150541&   149016& 148620  &           \\
 $5p^54f\ ^3G_4$&  156371 & 145174& 144421&  151285&   149622& 149210  &           \\
 $5p^54f\ ^3D_3$&  157255 & 145363& 144693&  150949&   149464& 149153  &  149927.1 \\
 $5p^54f\ ^3G_3$&  161005 & 149159& 148330&  155332&   153519& 153130  &  153339.1\\
 $5p^54f\ ^3F_4$&  164624 & 153238& 152229&  159590&   157645& 157252  &           \\
 $5p^54f\ ^3F_2$&  168059 & 157135& 155922&  163326&   161024& 160592  &  160486.4\\
 $5p^54f\ ^3G_3$&  177237 & 165352& 164376&  171959&   169935& 169562  & \\
 $5p^54f\ ^3F_3$&  180229 & 168297& 167235&  174850&   172780& 172445  &
 \end{tabular}
\end{ruledtabular}
\end{table*}

The electronic configuration of the Th$^{4+}$ and U$^{6+}$ excited
states makes accurate calculations difficult: the ground state
configuration is a Rn-like closed shell system [Hg]$6p^6$,  while the first two
excited states have a hole in the $6p$ shell, resulting in the $6p^5
5f$ configuration. Since both of these configurations are of even
parity, they have to be included in the calculations on the same
footing, i.e. including the mixing of these configurations. In this
work, we separate the treatments of the electronic correlations into
two problems: (1) treatment of strong valence-valence correlations
and (2) inclusion of core excitations from the entire core.
We test the predictive ability of our method on the homologue case
of Xe-like La$^{3+}$ where the energies have been measured to high
precision.

\section{Method}

\begin{table*} [ht]
\caption{\label{tab2} Energies of the Th$^{4+}$ and U$^{6+}$ even
states calculated using the CI+all-order method. The results
obtained considering $6s^2$ to be valence electrons are listed in
rows labelled ``8-el''. Results obtains with small and large sets of
configuration are given in the corresponding rows. }
\begin{ruledtabular}
\begin{tabular}{lrrrclrrr}
    \multicolumn{4}{c}{Th$^{4+}$} &
   \multicolumn{1}{c}{} &
     \multicolumn{4}{c}{U$^{6
     +}$} \\
 \multicolumn{1}{c}{Level} &
\multicolumn{1}{c}{COWAN} &
\multicolumn{1}{c}{8el-small} &
\multicolumn{1}{c}{8el-large} &
 \multicolumn{1}{c}{} &
 \multicolumn{1}{c}{Level} &
\multicolumn{1}{c}{COWAN} &
\multicolumn{1}{c}{8el-small} &
\multicolumn{1}{c}{8el-large} \\
\hline \\[-0.8pc]
$6p^6  \ ^1S_0$&        0 &      0&       0 &  &$6p^6  \ ^1S_0$ &       0 &       0&       0 \\
$6p^55f\ ^3D_1$&  135013  & 137121&  134995 &  &$6p^55f\ ^3D_1$ &   87975 &   92458&   90850 \\
$6p^55f\ ^3D_2$&  140469  & 142088&  139842 &  &$6p^55f\ ^3D_2$ &   94775 &   98554&   96863 \\
$6p^55f\ ^3G_4$&  143819  & 145579&  143160 &  &$6p^55f\ ^3G_4$ &   97064 &  101288&   99539 \\
$6p^55f\ ^3G_5$&  145606  & 147001&  144714 &  &$6p^55f\ ^3F_3$ &  102529 &  105492&  103627 \\
$6p^55f\ ^3F_3$&  147698  & 148752&  146314 &  &$6p^55f\ ^3G_5$ &  101946 &  105732&  104079 \\
$6p^55f\ ^1F_3$&  150769  & 150615&  148081 &  &$6p^55f\ ^1F_3$ &  107300 &  109051&  107240 \\
$6p^55f\ ^3F_4$&  156377  & 157221&  154552 &  &$6p^55f\ ^3F_4$ &  113656 &  116417&  114499 \\
$6p^55f\ ^1D_2$&  160980  & 162300&  159248 &  &$6p^55f\ ^1D_2$ &  116277 &  119989&  117725 \\
$6p^55f\ ^3G_3$&  209865  & 208535&  205597 &  &$6p^55f\ ^3G_3$ &  188185 &  186470&  183699 \\
$6p^55f\ ^3D_3$&  215174  & 213424&  210417 &  &$6p^55f\ ^3D_3$ &  196853 &  193827&  190953 \\
$6p^55f\ ^3G_4$&  217527  & 217054&  214015 &  &$6p^55f\ ^3F_2$ &  196653 &  197749&  193197
\end{tabular}
\end{ruledtabular}
\end{table*}

We use a hybrid approach developed in
\cite{Koz04,SafKozJoh09,PKST16} that efficiently treats these two
problems  by combining configuration interaction (CI) and a
linearized coupled-cluster methods, referred to as the CI+all-order
method. The first problem is treated by a large-scale CI method in
the valence space. The many-electron wave function is obtained as a
linear combination of all distinct many-electron states of a given
angular momentum $J$ and parity:
 \begin{align}
 \Psi_{J} = \sum_{i} c_{i} \Phi_i\,.
 \end{align}
Usually, the energies and wave functions of the low-lying states are
determined by diagonalizing the Hamiltonian in the CI method:
 \begin{align}
 H=H_{1} + H_2\,,
 \end{align}
where $H_{1}$ is the one-body part of the Hamiltonian, and $H_2$
represents the two-body part, which contains Coulomb + Breit matrix
elements. In the CI+all-order approach this bare Hamiltonian is
replaced by the effective one,
\begin{eqnarray}
H_1 &\rightarrow &H_1+\Sigma_1\,,\\
H_2 &\rightarrow& H_2+\Sigma_2\,,
\end{eqnarray}
where $\Sigma_i$ corrections incorporate all single and double
excitations from \textit{all} core shells to all basis set orbitals
(up to $n_{\textrm{max}}=35$ and $l_{\textrm{max}}=5$, efficiently
solving the second problem. The effective Hamiltonian $H^\text{eff}$
is constructed using a coupled cluster method \cite{BJLS89}. The
size of the core rather weakly affect the accuracy of the
CI+all-order approach for $Z\gtrsim 20$ and the method was used even
for superheavy atoms with $Z>100$.

\section{Method tests - {L\lowercase{a}$^{3+}$} calculation}

To test the method, we carried out the calculation for a homologue
system, La$^{3+}$, which has [Pd]$5s^2 5p^6$ ground state and $5s^2 5p^5 4f$
low-lying configurations.   The experimental values for relevant
La$^{3+}$ states are available \cite{NIST} for benchmark
comparisons. We started with the assumption that $5s^2$ shell may be
kept closed.  In this calculation, we  used a $V^{N-6}$ Dirac-Hartree-Fock (DHF) starting potential \cite{Dzu05}, where $N$ is the number of electron, i.e.
the potential of Cd-like ionic core of La$^{9+}$. Such calculation
 yielded poor results for the excited states of
interest. Further tests showed that the $5s6s5p^54f$ configuration
gives the largest contribution to the low-lying states after  the
$5s^2 5p^6$ and $5s^2 5p^5 4f$ configurations. The next  largest
contributions come from the $5s^2 5p^5 5f$ and $5s^2 5p^5 6p$
configurations, as expected. Therefore, the La$^{3+}$ calculations
have to be carried out as a 8-valence electron computation, with
both $5s$ and $5p$ shells open. We  used a $V^{N-8}$ starting potential,  i.e.
the potential of the  La$^{11+}$ ionic core. To construct the set of the most
important even parity configurations, we started ionic core with the $5s^2
5p^6$ and $5s^2 5p^5 4f$ configurations and allowed to excite one or
two electrons from these configurations to excited states up to
$7f$. This produced the list of 3277 (relativistic) configurations
resulting in 360~633 Slater determinants. Below we refer to this run
as ``small''. For the next run, we reordered the original set of
configurations by their weight and allow further one-two excitations
up to $7f$ electrons from the 21 configurations with highest
weights. This (medium) set has 11785 configurations and 3~453~220
determinants, making it ten times larger than the small run. Note
that it is the number of Slater determinants that defines the
computational time. Finally, we also allow a single excitation from
the 59 highest weight configurations to all  electrons up
$20spd16f12g$. This (large) run has 18187 configurations and
4~187~914 determinants.

The results for the energies of even states of xenon-like lanthanum
(La$^{3+}$) are summarized in Table~\ref{tab1}. The CI+all-order
results obtained considering $5s^2$ to be a core shell are listed in
rows labeled ``6-el''. The CI+all-order results obtained considering
$5s^2$ to be valence electrons are listed in rows labeled ``8-el''.
Results obtains with small, medium and large sets of configuration
are given in the corresponding rows.  The results are compared with
experimental data from NIST database \cite{NIST}. The COWAN code \cite{cowan}
data are given for reference. The table clearly demonstrates
problems of the 6-el approach. The differences between medium and
large runs all relatively small. The results of the small run are
larger than the experimental values and the results of both larger
runs are smaller than the experimental values. Therefore,  the
accuracy of the inclusion of the core-valence correlations via the
effective Hamiltonian is comparable with the contribution of the
remaining configurations. Since the inclusion of further
configurations can only lower the values, inclusion of the further
configuration will not improve the accuracy of the theory.

\begin{table*} [ht]
\caption{\label{tab3} Transition energies (in cm$^{-1}$), matrix
elements M1 (in $\mu_B$) and E2 (in $ea_0^2$), transition rates (in
$1/s$), and lifetimes (in $s$) of the $ 6p^55f\ ^3D_{1,2}$ levels in
Th$^{4+}$ and U$^{6+}$. }
\begin{ruledtabular}
\begin{tabular}{lllcccccc}
\multicolumn{1}{c}{Ion}&
\multicolumn{1}{c}{Upper}&
\multicolumn{1}{c}{Lower}&
\multicolumn{1}{c}{}&
\multicolumn{1}{c}{Transition} &
\multicolumn{1}{c}{Matrix}&
\multicolumn{1}{c}{Transition}&
\multicolumn{1}{c}{Branching}&
\multicolumn{
1}{c}{Lifetime}\\
\multicolumn{1}{c}{}&
\multicolumn{1}{c}{level}&
\multicolumn{1}{c}{level}&
\multicolumn{1}{c}{}&
\multicolumn{1}{c}{energy} &
\multicolumn{1}{c}{element}&
\multicolumn{1}{c}{rate}&
\multicolumn{1}{c}{ratio}&
\multicolumn{1}{c}{}\\
\hline\\[-0.8pc]
U$^{6+}$  &$ 6p^55f\ ^3D_1$& $6s^2\ ^1S_0$&  M1&    90850   &5.8E-04&   0.0023  & 1    &450    \\
Th$^{4+}$ &$ 6p^55f\ ^3D_1$& $6s^2\ ^1S_0$&  M1&    134994  &5.8E-04&   0.0070  & 1    &130   \\   [0.4pc]
U$^{6+}$  &$ 6p^55f\ ^3D_2$& $6s^2\ ^1S_0$&  E2&    96863   &0.18278 &  6.38    &0.610 &    \\
          &$ 6p^55f\ ^3D_2$& $6p^55f\ ^3D_1$&  M1&  6013    &1.86646 &  4.09    &0.390 &  \\
          &$ 6p^55f\ ^3D_2$& $6p^55f\ ^3D_1$&  E2&  6013    &-0.039&    2.7E-07 &0.000 & 0.09554 \\  [0.4pc]
Th$^{4+}$ &$ 6p^55f\ ^3D_2$& $6s^2\ ^1S_0$&  E2&    139842  &0.4786  &  274     &0.992 &      \\
          &$ 6p^55f\ ^3D_2$& $6p^55f\ ^3D_1$&  M1&  4848    &1.9126 &   2.25    & 0.008&    \\
          &$ 6p^55f\ ^3D_2$& $6p^55f\ ^3D_1$&  E2&  4848    &-0.320 &  6.2E-06  & 0.000& 0.00361 \\  [0.4pc]
\end{tabular}
\end{ruledtabular}
\end{table*}

 \begin{table} [ht]
\caption{\label{tab4} $E2$ $6p^55f\ ^3D_{2} - 6p^55f\ ^3D_{1}$ reduced matrix element in $ea_0^2$.}
\begin{ruledtabular}
\begin{tabular}{llcc}
\multicolumn{1}{c}{}&
\multicolumn{1}{c}{}&
\multicolumn{1}{c}{Th$^{4+}$}&
\multicolumn{1}{c}{U$^{6+}$} \\
\hline \hline \\[-0.8pc]
    no RPA& Small&  0.348 &    0.0399   \\
    RPA   &  Small& 0.301 &     0.0026 \\
    RPA  &  Medium& 0.303  &   0.0003  \\
    no RPA& Large  &       &    0.0389 \\
    RPA    & Large& 0.320   &  0.0034   \\
\end{tabular}
\end{ruledtabular}
\end{table}

\section{{T\lowercase{h}}$^{4+}$ and U$^{6+}$ CI+all-order calculations}

We use the results of La$^{3+}$ tests to construct the Th$^{4+}$ and
U$^{6+}$ configuration sets for 8 valence electrons with about
3~800~000 determinants.
We  used a $V^{N-8}$ starting potentials,  i.e.
the potentials of the  Th$^{12+}$ and U$^{14+}$  ionic cores.
The resulting energies are given in
Table~\ref{tab2}.
Results obtains with small and large sets of configurations are
given in the corresponding rows. The small set is equivalent to the
La$^{3+}$ small set. The results for the reduced $M1$ (in $\mu_B$)
and $E2$ (in $ea_0^2$) matrix elements between first three states
are given in Table~\ref{tab3}. Corresponding transition energies in
cm$^{-1}$, transition rates in s$^{-1}$, branching ratios, and
lifetimes of the $6p^55f\ ^3D_{1,2}$ states in seconds are also
listed. The transition rates (in s$^{-1}$) are obtained as
\begin{eqnarray}
  A(M1) = \frac{2.69735\times 10^{13}}{(2J+1)\lambda ^{3}}  S(M1)\,,\\
   A(E2) = \frac{1.11995\times 10^{18}}{(2J+1)\lambda ^{5}} S(E2)\,,
\end{eqnarray}
where $S$ is the square of the reduced matrix elements (in $\mu_B$
for $M1$ and in $ea_0^2$ for $E2$), $\lambda$ is the transition
wavelength in \AA, and $J$ is the total angular momentum of the
upper state. The branching ratios are the ratios of the rate of the
given transition to the total rate, and the lifetime is the inverse
of the total transition rate.

All of the values are obtained from the ``large'' runs. We found
only a weak dependence of the matrix elements on the size of the
configuration space with the exception of the $6p^55f\ ^3D_{2} -
6p^55f\ ^3D_{1}$ $E2$ matrix element in U$^{6+}$. The values of this
$E2$ matrix element in Th$^{4+}$ and U$^{6+}$ calculated with small,
medium, and large number of configurations are listed in
Table~\ref{tab4}. The final results in Table~\ref{tab3} all include
the random-phase approximation (RPA) correction to the M1 and $E2$
operators. In Table~\ref{tab4} we listed the values without the RPA
corrections as well. It is clear that while the value of this matrix
element are similar for all runs in Th$^{4+}$, this is not the case
in U$^{6+}$. This is explained as follows. The dominant one-electron
contributions to the $6p^55f\ ^3D_{2} - 6p^55f\ ^3D_{1}$ $E2$ matrix
element come from the $6p_{3/2}-6p_{3/2}$ and $5f_{5/2}-5f_{5/2}$,
and $5f_{5/2}-5f_{7/2}$ matrix elements. These contributions
strongly cancel, leading to a small final value. In Th$^{4+}$ there
is noticeable addition from the configurations containing $7p$ and
$6f$ orbitals, which are mixed with the $6p^55f$ configuration.

In U$^{6+}$, the $5f$ electron becomes stronger bound and closer to
the ground configuration. This is due to a level crossing mechanism,
which is responsible for the presence of optical transitions in
highly charged ions \cite{BerDzuFla10}. As a result, the
configuration mixing with higher orbitals, such as $7p$ and $6f$, is
suppressed. For example, the admixture of the $6f$ orbital to the
$6p^55f$ configuration is almost two times larger in Th$^{4+}$, than
in U$^{6+}$. This weakens the cancelation between $np-np$ and
$nf-nf$ contributions. As a result, the $E2$ matrix element $6p^55f\
^3D_{2} - 6p^55f\ ^3D_{1}$ in U$^{6+}$ is highly dependent on the
details of the calculations, but is more stable for Th$^{4+}$.
Because of that for uranium we can not predict this amplitude
reliably. It is certainly significantly smaller than in Th$^{4+}$,
most likely by an order of magnitude.

\section{Summary of the differences between the {T\lowercase{h}}$^{4+}$ and U$^{6+}$ results for low-lying levels}

Below, we outline the resulting differences between the low-lying
metastable levels  of the two ions.
\begin{itemize}
 \item The $6p^5 5f$ levels lie \textit{closer} to the ground state in
U$^{6+}$ than in Th$^{4+}$. This is expected, as in hydrogenic ions
the $5f$ shell lies below the $6p$ shell. Therefore, the level
crossing must take place along the isoelectronic sequence.
 \item The lifetime of the first, $^3D_1$, excited state in U$^{6+}$ is
more than 3 times longer (450 s). This is purely due to smaller
transition energy in U$^{6+}$  as the M1 matrix element is
practically the same. Note that $M1$ transition rates scale as
$\lambda^{-3}$.
 \item The lifetime of the second, $^3D_2$, excited state in U$^{6+}$ is 26
times longer. This is both due to $\lambda^{-5}$ scaling of the $E2$
transition rate and smaller $E2$ matrix element, compared to
Th$^{4+}$.
 \item The branching ratio of the $^3D_2$ level to the
ground state is over 99\% in Th ion, but only 61\% in U ion. As a
result, the large fraction of the U ions from $^3D_2$ state ends up
in a highly metastable $^3D_1$ level, but very few Th ions do.
 \item As described above, the $^3D_2 - {}^3D_1 $ $E2$ matrix element is
much smaller in U ion. We note that this $E2$ transition is
extremely weak in both cases and $M1$ decay is orders of magnitude
stronger. The $M1$ matrix elements $^3D_2 - {}^3D_1$ are similar in
both ions and $M1$ transition rate is factor of 2 larger in U ion.
\end{itemize}

\section{Conclusion}

We have established that the $ns$ electrons have to be considered as
valence for an accurate determination of the properties of
particle-hole states with a hole in the respective $np$ shell. We find that the CI+all-order method
works well with the  $V^{N-8}$ 
starting potential which extends the applicability of this approach.
We have developed an algorithm for the efficient construction of the
large-scale CI  configuration sets. The methodology is tested on the
La$^{3+}$ ion and excellent agreement with experiment is obtained.
These results suggest that the uncertainties of our predictions for
the energy levels in Th$^{4+}$ and U$^{6+}$ ions are expected to be
less than 1\%. Matrix elements, branching ratios and lifetimes of
the Th$^{4+}$ and U$^{6+}$ low-lying states were calculated and
analyzed.
\section*{Acknowledgement}

We thank Steve Lundeen for bringing our attention to the problem of
RESIS measurements in U$^{6+}$ and helpful discussions. This work is
partly supported by NSF Grant No. PHY-1620687. M.G.K.  acknowledges
support from Russian Foundation for Basic Research under Grant No.
17-02-00216.

%

\end{document}